\documentclass[useAMS,usenatbib]{mn2e}
\usepackage{graphics,graphicx,epsfig,psfig}
\usepackage[normalem]{ulem}
\usepackage[dvipsnames]{color}
\usepackage{soul,xcolor}
\usepackage{amsmath, amssymb}
\usepackage[]{inputenc,amssymb}
\usepackage{booktabs,caption}
\usepackage{amsmath}
\usepackage{multirow}
\usepackage{soul}

\setstcolor{red}

\def \be{\begin{equation}}
\def \ee{\end{equation}}
\def \ba{\begin{eqnarray}}
\def \ea{\end{eqnarray}}
\def \etal{{et al.}}

\definecolor{webgreen}{rgb}{0,.5,0}
\definecolor{webbrown}{rgb}{.6,0,0}

\usepackage[%
    colorlinks = true,%
    linkcolor = blue,%
    urlcolor  = blue,%
    citecolor = webgreen,%
    anchorcolor = blue]{hyperref}

\newcommand{\ufhref}[3][blue]{\href{#2}{\color{#1}{#3}}}%

\title[Cosmic ray heating of IGM]{Cosmic ray heating of intergalactic medium: patchy or uniform?}
\author[Jana, Nath]
{Ranita Jana, Biman B. Nath\\
Raman Research Institute, Sadashiva Nagar, Bangalore 560080, India
}
\begin{document}
\maketitle
\label{firstpage}
\begin{abstract}
We study the heating of the intergalactic medium (IGM) surrounding high redshift star forming galaxies due to cosmic rays (CR). We take into account the diffusion of low energy cosmic rays and study the patchiness of the resulting heating. We discuss the case of IGM heating around a high redshift minihalo ($z\sim 10\hbox{--}20$, M$\sim 10^5\hbox{--}10^7$ M$_\odot$),and put an upper limit on the diffusion coefficient $D\le 1\times 10^{26}$ cm$^2$ s$^{-1}$ for the heating to be inhomogeneous at $z\sim 10$ and $D\le 5\hbox{--}6 \times 10^{26}$ cm$^2$ s$^{-1}$ at $z\sim 20$. For typical values of $D$, our results suggest uniform heating by CR at high redshift, although there are uncertainties in magnetic field and other CR parameters. We also discuss two cases with continuous star formation, one in which the star formation rate (SFR) of a galaxy is high enough to make the IGM in the vicinity photoionized, and another in which the SFR is low enough to keep it neutral but high enough to cause significant heating by cosmic ray protons.
In the neutral case (low SFR), we find that the resulting heating can make the gas hotter than the cosmic microwave background (CMB) radiation for $D < 10^{30}$ cm$^2$ s$^{-1}$, within a few kpc of the galaxy, and unlikely to be probed by near future radio observations. In the case of photoionized IGM (high SFR), 
the resulting heating of the gas in the vicinity of high redshift ($z\sim 4)$ galaxies of mass $\ge 10^{12}$ M$_\odot$ can suppress gas infall into the galaxy. At lower redshifts ($z\sim 0$), an SFR of $\sim 1$ M$_\odot$ yr$^{-1}$ can suppress the infall into galaxies of mass $\le 10^{10}$ M$_\odot$.
\end{abstract}
\begin{keywords} Galaxies: evolution, intergalactic medium -- cosmic rays -- supernovae: general -- dark ages, reionization, first stars
\end{keywords}
\section{Introduction}
\label{sec:intro}
Galaxies interact with the surrounding gas in various ways, through gravitational and mechanical means, as well as through radiation. The gravitational
field of collapsed structures help them to accrete matter from  surrounding regions, setting up an inflow of gas and dark
matter. The radiation emanating from stars and possible active galactic nuclei (AGN) also affect the intergalactic medium (IGM),
by ionizing and heating. The process of star formation and AGN activities stir up the interstellar medium (ISM) of the galaxies,
often setting up galactic outflows which interact with the IGM gas through fluid dynamical interactions. There is yet another 
type of interaction that has been discussed in the literature,  through high energy particles, which may be produced during the
star formation or AGN activity in galaxies.

\cite{Ginzburg1966} pointed out that cosmic rays (CR) accelerated in supernovae (SNe) and radio galaxies could raise the temperature
of the IGM gas to $\ge 10^5$ K. Their argument was based on the fact that low energy CRs lose a large fraction
of their energy through Coulomb interactions.  \cite{Nath1993} addressed the question of possible reionization of the Universe through
such processes, 
and concluded that it would require a very large star formation rate (SFR) density. 
According to \cite{Lacki2015}, CRs  would
contribute towards a significant non-thermal pressure of the IGM gas. Another possible effect of CRs in the IGM discussed in the 
literature is the production of $^6 Li$ by CR $\alpha$ particles.  \cite{Nath2006} showed the observed abundance ratio of $^6 Li/H$ can 
be related to the observed entropy of the intracluster medium, through Coulomb heating.

Recently, \cite{Sazonov2015} suggested that
 low energy CRs  (with kinetic energy $\le 30$ MeV per nucleon) could have 
heated the neutral IGM at high redshift and  change the HI emission characteristics of the gas. Such a feature can be
potentially detected in  planned experiments that will detect redshifted 21 cm emission. Following this argument,
\cite{Leite2017} calculated in detail the heating of the IGM by CRs, and found that the IGM temperature could have 
been increased by $\Delta T \sim 100$ K at $z\sim10$.

However, as \cite{Leite2017} and others have pointed out, the propagation of the CRs crucially depend on the diffusion coefficient. Diffusion of CRs can heat the surrounding gas in a non-uniform manner.
 In this paper, we discuss the heating of intergalactic medium by CRs produced by star forming galaxies, as a function
of diffusion coefficient,gas density, SFR and other relevant parameters, and discuss the implication of this kind of heating.

\section{Preliminaries}
We consider the heating effect of CR protons in this paper on the IGM gas surrounding a galaxy. The energy deposition by protons
depends on several parameters, and below we list our assumptions regarding them. 

\subsection{CR spectrum}
The CR luminosity of a galaxy is determined by its SFR, assuming a Salpeter IMF (with $0.1$ M$_\odot$ and $30$ M$_\odot$ as the lower and upper limits), with a total mechanical energy output of $10^{51}$ erg per SNe and an efficiency of $\eta $ for CR acceleration. 
There is a significant uncertainty in this parameter (from less than 0.1 to $\sim$ 0.5).
 We assume a value of $\eta=0.1$, which is supported  by the simulation results of \cite{caprioli2014}.
This gives us,
\be
L_{\rm cr} =2 \times 10^{40} \, {\rm erg} \, {\rm s}^{-1} \, \Bigl ({\eta \over 0.1} \Bigr ) \, \Bigl ( { {\rm SFR} \over 1 \, {\rm M}_\odot \, {\rm yr}^{-1}} \Bigr ) \,.
\label{norm}
\ee
We assume that the CR protons leave the
galactic virial radius $R_{vir}$, with a spectrum $n_{cr} (p_0) \propto p_0^{\alpha}$, where $p_0 (\equiv p(r=R_{vir}))$ denotes the momentum of protons as they leave the galaxy at $r=R_{vir}$, and $n_{cr} (p_0)$ is the rate of CRs (number of CR per second) coming out of the galaxy 
with momentum $p_0$. 
In our calculation, we assume $\alpha=-2.5$, and discuss the effect of changing its value later in the paper.

The spectrum is normalized such that the total energy flux corresponds to the above mentioned CR luminosity, or,
\be
L_{cr}=\int_{p_{0,min}}^{p_{0,max}} E_k n_{cr}(p_0) dp_0 \,,
\label{norm2}
\ee
where $p_{0,min}$ and $p_{0,max}$ are the lower and upper limits of CR momenta, with kinetic energy $E_k=\sqrt{p^2c^2 + m_p^2 c^4}-m_pc^2$.
By denoting $x_0=p_0/(m_pc)$,
 CR spectrum can be written (in terms of number of particles per unit time) as,
\be
n_{cr}(p_0) dp_0=\Bigl ( {L_{\rm cr} \over (m_pc^2) (m_pc)^{\alpha+1} \int [\sqrt{1+x_0^2} -1]x_0^\alpha dx_0}\Bigr ) p_0^{\alpha} dp_0 \,,
\label{eq:sp1}
\ee
where the integration is carried out between $x_{0,min}={p_{0,min} \over m_p c}$ and $x_{0,max}= {p_{0,max} \over m_p c}$.
We will find it more convenient to describe the spectrum in terms of $\beta(\equiv v/c)$, and in the rest of the paper, we will
write the emergent spectrum from the galaxy in equation \ref{eq:sp1} as $n_{cr}(\beta_0)d\beta_0$, where $p_0={m_p \beta_0 c \over \sqrt{1-\beta_0^2}}$.

The heating effect of CRs is insensitive to the upper limit of energy, but depends strongly on the lower limit, since the energy loss
rate increases with decreasing CR energy. Previous works by \cite{Sazonov2015} and \cite{Leite2017} considered heating by protons
of  $\le 10\hbox{--}30$ MeV. However, protons with very low energy, with $\le 1$ MeV, are unlikely to survive the interactions with the interstellar medium (ISM)
of the parent galaxy for the following reasons. The loss of (kinetic) energy by a proton through interaction with ionized gas of electron density
$n_e$ cm$^{-3}$ and temperature $T_e$ K is given by \citep{Mannheim1994}
(their equation 4.22)
\be
-{d E_k \over dt} \approx 5 \times 10^{-19} \, {{\rm erg} \over {\rm s}} \Bigl ( {n_e \over {\rm cm}^{-3}} \Bigr ) {\beta ^2 \over x_m^3 +\beta ^3} \,, 
\label{mannheim}
\ee
where $x_m=0.0286 (T_e /2 \times 10^6 \, {\rm K})^{1/2}$.\\
For a CR proton in a neutral medium with particle density $n_{\rm HI}$, the energy loss rate is given by equation (4.32) of \cite{Mannheim1994},
\ba
 -\Big(\frac{dE_k}{dt}\Big)&=&3\times 10^{-19} \, {{\rm erg} \over {\rm s}} \, \Big({\frac{n_{\rm HI}}{cm^{-3}}}\Big)
 \nonumber\\ && \times (1+0.0185 \, \ln \beta  \, H[\beta-\beta_c]) 
  \frac{2 \beta^2}{\beta_c^3+2 \beta^3} \,,
  \label{eq:energyloss-neutral}
\ea
where $\beta_c\approx 0.01$,  corresponding to the orbital speed of electrons in a hydrogen atom. For analytical simplicity we neglect
the term $(1+0.0185 \,  \ln \beta  \, H[\beta-\beta_c])$ in the above expression, since it  does not significantly affect the result.

\begin{figure}
\centering
\includegraphics[height= 2in]{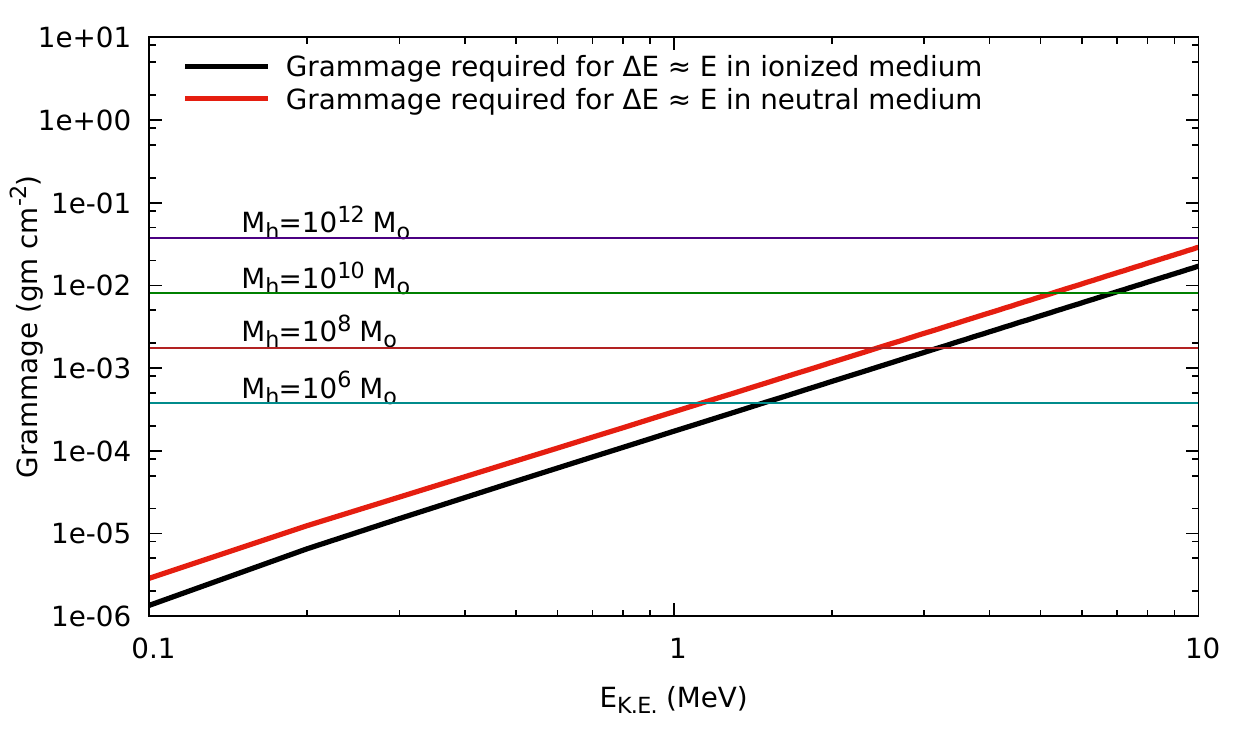}
\fontsize{10}{10}
\caption{Grammage to deplete the total energy of a proton is shown against the proton kinetic energy  for ionized and neutral media. Corresponding
line-of-sight grammages for galaxies with  $\rm M_h=10^6,10^8, 10^{10}, 10^{12}$ M$_\odot$ are shown for  $z=10$. }
\label{fig:grammage}
\end{figure}

   This can be used to estimate the grammage ($\int n_e m_p \beta c dt$) required to drain
a proton of its kinetic energy (when $E_k \approx dE_k$). We show this value of grammage as a function of proton kinetic energy in Figure \ref{fig:grammage}.
We also superimpose several relevant values of line-of-sight grammage as horizontal lines, corresponding to different galaxy masses  at redshift $10$,
estimated as $M_h f_b \over R_{vir}(z) ^2$, where $M_h$ is the total halo mass of a galaxy, $f_b\approx 0.15$ is the cosmic baryon fraction and
$R_{vir} (z)$ is the virial radius at redshift $z$.

The curves in the figure show that protons of energy less than $ 0.1\hbox{--} 1$ MeV are likely to lose all their energy if they were to travel
through the ISM of the parent galaxy in a straight path. Therefore, this value should indicate the lower limit of energy, $E_{\rm min}$,
corresponding to the lower limit of momentum $p_{0,min}$ mentioned above.

However, there are two possibilities that can change the lower limit.
CRs may {\it diffuse} through the ISM and the corresponding grammage is likely
to be much higher than depicted as horizontal lines here. For example, in the case of Milky Way,  the inferred grammage
of $\sim 10$ g cm$^{-2}$ is much larger than the total column density of the disk, and the corresponding $E_{\rm min}
\sim 50$ MeV, where a break in CR spectrum is expected and is indeed observed \citep{Nath2012}.

On the other hand,  
CRs in star forming galaxies may be {\it advected} 
by gas in the ensuing galactic outflow. In this case, the lower limit of energy may decrease because of adiabatic loss. Since
the adiabatic loss of CR  energy scales as $\Delta E \propto R^{-1}$,
the factor by which the lower limit of energy will decrease is $\epsilon_{\rm adv}$ roughly the ratio of the size of a galaxy halo
and the region of CR production, {\it i.e.}, the disk of a star forming galaxy \citep{Nath1993}. This ratio is roughly of order $\sim 0.1$,
and we adopt this value for $\epsilon_{\rm adv}$.

Therefore, we assume two values of the lower limit of energy, $0.1$ and $1$ MeV to encapsulate the uncertainties in the 
processes of CR propagation until it reaches the IGM.

\subsection{CR diffusion}
For heating of the gas in which CRs propagate, it is the low energy CRs that play an important role, since the amount of energy 
lost by a CR increases with decreasing energy. In our context, the energy range of interest is $\le 100$ MeV, as has been pointed out by 
previous workers \citep{Sazonov2015, Leite2017}. 

The diffusion coefficient of CRs is believed to depend on energy, and this dependence in the Milky Way is estimated from a comparison
of the observed CR spectrum with what is believed to be the source spectrum. There is, however, considerable uncertainty in the interpretation. Phenomenologically, a few prescriptions for the diffusion
coefficient are used in the literature and in models such as GALPROP (see, eg, \cite{ptuskin2012}). In the `plain diffusion model', the diffusion coefficient
for low energy cosmic rays is thought to be, $
D= 2.2 \times 10^{28} \beta^{-2} \, {\rm cm^2/s} 
$.
As has been pointed out by \cite{ptuskin2012}, the increase in the diffusion coefficient with decreasing
energy has no physical explanation and is purely a phenomenological inference. The other common model of diffusion coefficient is that
of `distributed reacceleration', which scales as $p^{1/3}$, and therefore decreases with decreasing energy (see Figure 1 of \cite{ptuskin2006}).
Therefore, it is not clear from phenomenological studies if the diffusion coefficient should increase or decrease with decreasing energy at low energies. 
For simplicity we assume the diffusion coefficient to be constant for low energy CR protons.

The diffusion coefficient also depends on the magnetic field, because the diffusion of CRs depend on 
particle scattering by magnetohydrodynamic (MHD) waves and irregularities. 
For Kolmogorov-type spectrum of turbulence, the diffusion coefficient for scattering of protons off magnetic irregularities
scales as $D \propto r_g^{2-5/3} \propto  B^{-1/3} 
$, where $r_g$ is the gyroradius \citep{ptuskin2012}.

In light of the above discussion, we assume the diffusion coefficient of low energy CRs to be constant, to be $2 \times 10^{28}$ cm$^2$ s$^{-1}$,
for a $5 \mu$G magnetic field in the Milky Way ISM. The present day IGM magnetic field strength is estimated to be of order $\sim 10^{-9}$ G
\citep{subramanian2016}. Using the expected
scaling of $D \propto B^{-1/3}$ for Kolmogorov spectrum of magnetic irreguarities, we estimate the diffusion coefficient at present epoch to be $D \sim 3 \times 10^{29}$ cm$^2$ s$^{-1}$. The magnetic field strength scales with redshift as $B \propto (1+z)^{2}$ since the magnetic energy density scales as $(1+z)^{4}$. This 
implies a value of the diffusion coefficient at $z\sim 10$ to be
$D \sim  10^{29}$ cm$^2$ s$^{-1}$. We use this as the fiducial value in our calculation for heating of the IGM gas at high redshift ($z\sim 10 \hbox{--}20$).
However, we discuss the effect of changing the diffusion coefficient on our results.
\subsection{Physical property of IGM near the galaxy}
We assume for simplicity a static and uniform density gas around the galaxy, with number density $n_{\rm IGM}$ cm$^{-3}$. The IGM density near a galaxy is likely to be larger than the critical matter density,  and we assume that it is a factor $\Delta \approx 10$ times the critical density at a given epoch. 
In other words,
\ba
\rho_{\rm IGM}(z) &=&10 \times f_b \rho_{cr} \Omega_m (z)
\nonumber\\
&=&  5.6 \times 10^{-27} \, {\rm cm}^{-3} \, 
\Big(\frac{\Delta}{10}\Big) \Bigl ( {1+z \over 11} \Bigr)^3\,.
\ea
The cosmological parameters used are determined by \citet{planck}.
Therefore our assumed number density of gas $(n_{\rm IGM}=\frac{\rho_{\rm IGM}}{\mu m_p})$ is $2.8 \times 10^{-3} \rm cm^{-3}$ in neutral medium ($\mu=1.2$) and $5.6 \times 10^{-3} \rm cm^{-3}$ in ionized medium ($\mu=0.6$) at redshift 10 whereas the number density of electron is $2.9 \times 10^{-3} \rm cm^{-3}$ since $\mu_e=1.14$.

We shall justify our assumption in \S 4.3 by showing that a uniform density with $\Delta =10$ approximately gives the same results as in the case of a density profile that is more realistic outside the halo, within the relevant distance.

The IGM near a galaxy is also likely to be photoionized, owing to the ionizing radiation from massive stars, even before the first SNe begin to produce CR. This can be demonstrated by estimating the Str\"omgren sphere radius for different values of SFR and $\Delta$. Using STARBURST99 we find that for a continuous star formation scenario,  the number of ionizing photons radiated per second is $\approx 2\times10^{54} ({\rm SFR} /10 \, {\rm M}_\odot \, {\rm yr}^{-1})  $ at $10$ Myr after the onset of star formation. This gives a Str\"omgren sphere radius as,
\ba 
R_s & \approx & 193.8 \, \, {\rm kpc}\, \Bigl({{\rm SFR} \over 10 \, {\rm M}_\odot \, {\rm yr}^{-1} } \Bigr)^{1/3} \Bigl ( {\Delta \over 10} \Bigr )^{-2/3} \Bigl(\frac {f_b}{0.157}\Bigr)^{-2/3} \nonumber\\ && \qquad \times \Bigl(\frac{h}{0.677}\Bigr)^{-4/3} \Big(\frac{\Omega_{m,0}}{0.309}\Big)^{-2/3} \Big(\frac{1+z}{11}\Big)^{-2}
\ea
\\
For $z=10$, the corresponding radius is $\sim $ 194 kpc $ ({\rm SFR} /10 \, {\rm M}_\odot \, {\rm yr}^{-1})^{1/3} $. This length scale is much larger than the diffusion length of CR particles with $D\sim 10^{29}$ cm$^2$ s$^{-1}$, for a time scale of $100$ Myr, the typical duration of a burst of star formation. In other words, the heating effect of CRs is limited to an ionized region, for galaxies with SFR $\sim 10$ M$_\odot$ yr$^{-1}$. For a smaller SFR, one may have to consider a neutral IGM.

\begin{figure}
\centering
\includegraphics[height= 1.75in]{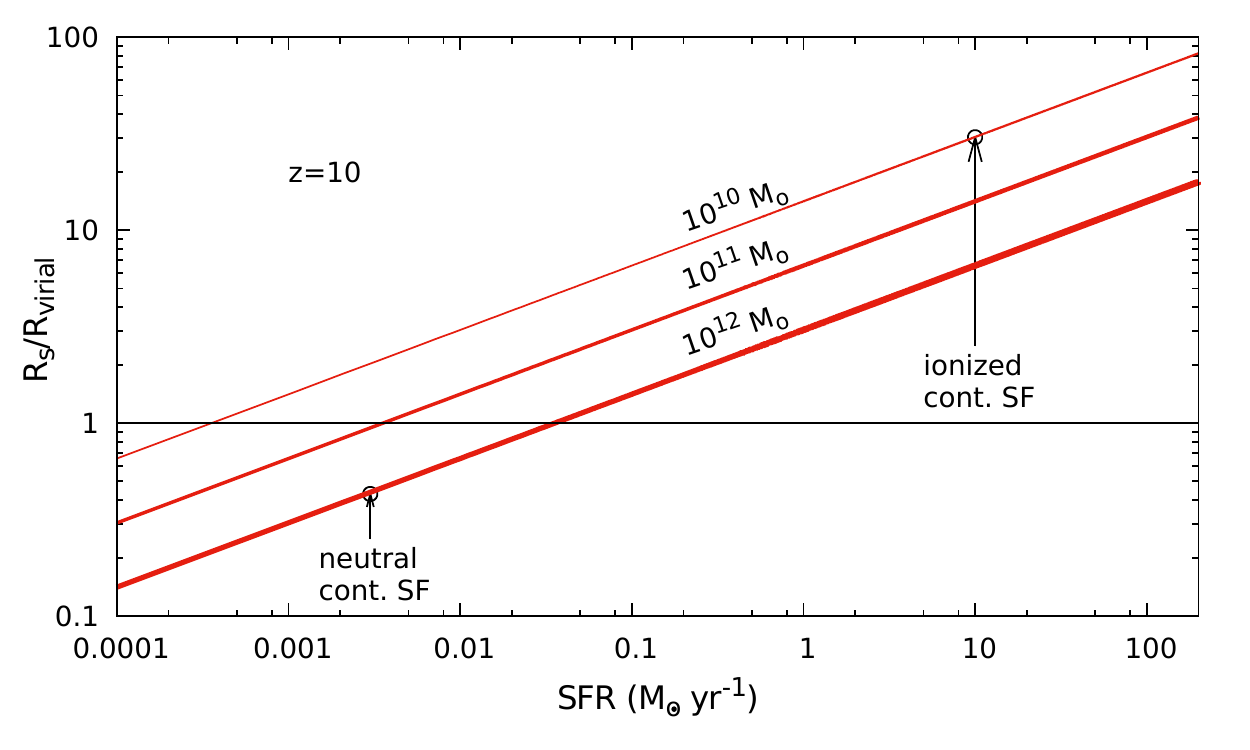}
\fontsize{13}{12}
\caption{Ratio of Str\"omgren Sphere radii and virial radii of galaxies of $M_h=10^{10}, 10^{11}, 10^{12}$ M$_\odot$ are shown as a function of SFR at $z=10$. Arrow marks show the fiducial SFR chosen for the cases of neutral and ionized IGM.}
\label{fig:virial-radius}
\end{figure}

In order to ascertain whether or not the IGM in the vicinity of the galaxy is neutral
or ionized,
the size of the ionized region can also be compared with the virial radius of a galaxy. We plot in Figure \ref{fig:virial-radius} the ratio of Str\"omgren radius to virial radius, as a function of SFR, for three different galaxy masses 
$M_h=10^{10}, 10^{11}, 10^{12}$ M$_\odot$ at $z=10$. The lower half region of the figure, with $R_s/R_{\rm vir} \le1$ refers to the case of neutral IGM gas
in the vicinity of a galaxy. For example, for a galaxy with mass $M_h=10^{10}$ M$_\odot$ at $z=10$, SFR has to be $\le 10^{-3}$ M$_\odot$ yr$^{-1}$ for the
gas to be neutral.

Therefore we consider two cases, one in which the surrounding gas is neutral and another in which it is photoionized. In the neutral case, we assume the
gas temperature to be $T=2.73 \times 151\times [(1+z)/151]^{2}=2.19 [(1+z)/11]^{2}$ K, since the matter and radiation temperature decouples at $z\sim 150$ and matter temperature
drops as $(1+z)^{-2}$ afterwards. In the photoionized case, we assume the gas to be at a temperature $10^4$ K, appropriate for a photoionzed gas
with primordial abundance.

In addition, we consider a third case, of that of a primordial supernova in a high redshift minihalo ($z=10\hbox{--}20$, M$_h=10^5\hbox{--}10^7$ M$_\odot$), which was discussed by \cite{Sazonov2015}.

\subsubsection{Gas cooling}
In the case of a photoionized IGM with primordial composition, we use the cooling due to bremsstrahlung and recombination cooling \citep{Efstathiou1992},
using the rates given in Appendix A by \cite{Hui1997}. 

For the case of neutral gas, the resulting temperature is small ($\le 1000$ K) and the cooling time exceeds the Hubble time. Therefore gas cooling
can be neglected in this case.

\section{Evolution of CR spectrum}
The calculation of the evolution of CR spectrum and resulting heating of the gas is described below for two different cases.

\subsection{Neutral IGM}

In case of neutral medium, we write \ref{eq:energyloss-neutral} in terms of $\beta$ of the proton,
\be
-\frac{d\beta}{dt}=3.9\times 10^{-16}\Bigl ({\frac{n_{\rm HI}}{cm^{-3}}}\Bigr ) \, \frac{\beta (1-\beta^2)^\frac{3}{2}}{\beta_c^3+2 \beta^3}
\ee

Furthermore, for a proton with diffusion coefficient $D$,
the rms speed is given by
\be
  \frac{dr}{dt}=
 \frac{3D}{r} \,.
 \label{diffusion}
\ee
Combining these two equations, we have,
\be
  {d\beta \over dr_{\rm kpc}} =1.2\times 10^{-2} \, \frac{(n_{\rm HI}/ {\rm cm}^{-3}) \, r_{kpc}}{(D/ 10^{29} {\rm cm}^2 \, {\rm s}^{-1})}\, \frac{\beta (1-\beta^2)^\frac{3}{2}}{\beta_c^3+2 \beta^3}\,,
\ee
where $r_{\rm kpc}$ is the distance from the virial radius of the galaxy in kpc unit.
This equation can be analytically solved to give $\beta$ as a function of $r$ given an initial value $\beta_0$.  This is given by,
 \ba
&& \tan(\arcsin \beta)-(\arcsin \beta)+(\beta_c^3/2) \ln \tan \Big(\frac{\arcsin \beta}{2}\Big)+\nonumber\\
&& \quad \frac{(\beta_c^3/2)}{\cos(\arcsin \beta)}\nonumber\\
&&=\tan(\arcsin \beta_0)-(\arcsin \beta_0)+(\beta_c^3/2) \ln \tan \Big(\frac{\arcsin \beta_0}{2}\Big)\nonumber\\
&& +\frac{\beta_c^3/2}{\cos(\arcsin \beta_0)}-{3.1\times 10^{-3}  {(n_{\rm HI}/{\rm cm}^{-3}) \, r_{\rm kpc}^2} 
\over (D / 10^{29} \, {\rm cm}^2 \, {\rm s}^{-1})} \,.
\label{dbetadr-neutral}
\ea
This relation can be used to trace the evolution of the CR spectrum as a function of distance $r$, given
a density of the medium.

 A useful parameter to define in this context is the distance through which the minimum energy CR proton
loses all its energy, which is calculated from equation  \ref{dbetadr-neutral}, by using a value of $\beta_0$ corresponding to $E_{0,min}$. 
This distance scale will be important in describing the results of temperature profile later in the paper.
We have found that this length scale $r_0$ can be approximately determined by,
\be
r_0 \approx 0.1\, {\rm kpc} \, \Bigl ( {E_{0,min} \over 1 \, {\rm MeV}} \Bigr )^{0.73} \, \Bigl ( {n_{\rm HI} \over 1\, {\rm cm}^{-3}} \Bigr )^{-0.54} \Big( \frac{D}{ 10^{29} \rm cm^2 s^{-1}}\Big)^\frac{1}{2}\,
\label{eq:rnot}
\ee
In the case of $z=10$ and $\Delta=10$, the corresponding particle density is $n_{\rm HI}\approx 0.003$ cm$^{-3}$, and a $1$ MeV proton
loses all its energy within a distance $\approx 2.2$ kpc from the virial radius. We will refer to these values when we discuss the effect of IGM heating.

\subsection{Photoionized IGM}
\begin{figure}
\centering
\includegraphics[height= 1.75in]{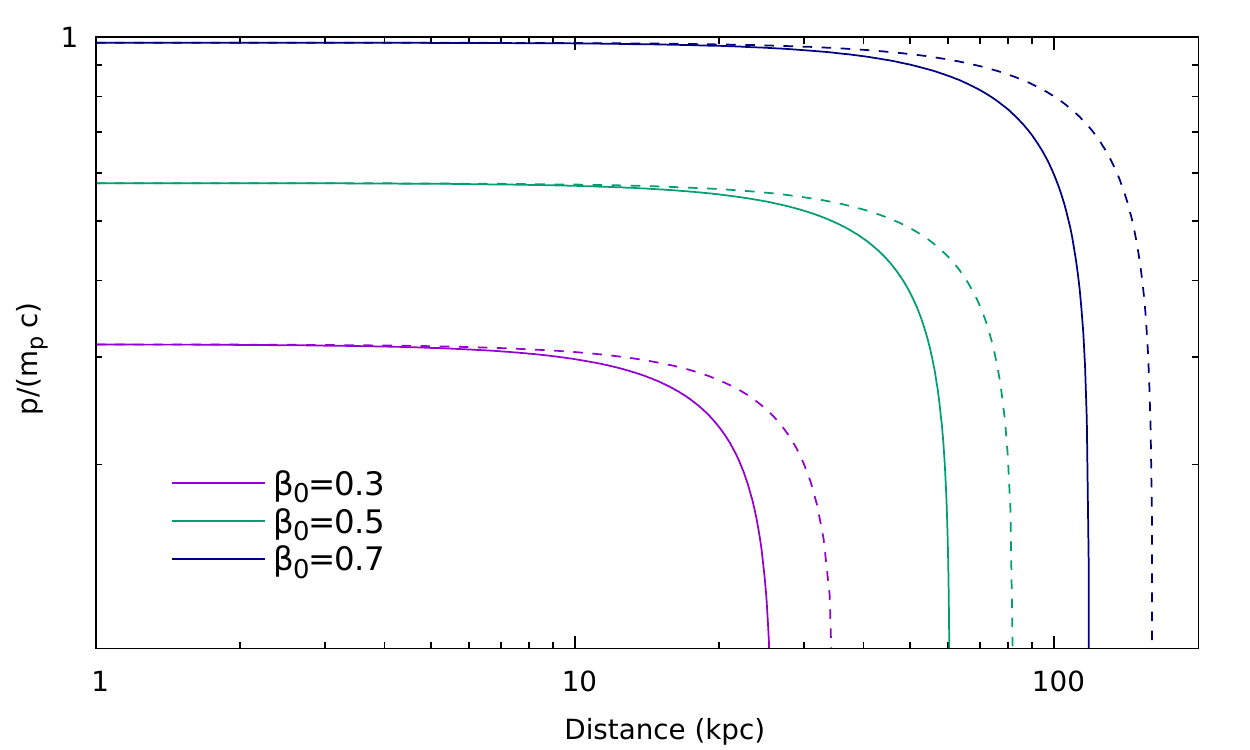}
\caption{Change in the dimensionless momentum $p/(m_p c)$ of a proton with distance for different initial values of $p$ (whose corresponding $\beta_0$ values are shown as labels), for  $z=10$ and $\Delta=10$. Solid lines show the
case of ionized IGM and dashed lines show the case of neutral IGM. }
\label{fig:momentum}
\end{figure}

In the case of photoionized gas, we can write equation \ref{mannheim} in terms of $\beta$, as,
\be
-\frac{d\beta}{dt}=3.3\times 10^{-16} \, {\rm s}^{-1} \, \Bigl ({n_e \over {\rm cm}^{-3}}\Bigr ) \, 
\frac{\beta (1-\beta^2)^\frac{3}{2}}{\beta^3+x_m^3} \,.
\ee
For analytical simplicity, we assume $x_m$ to be a constant and fix its value appropriate for
$T=10^4$ K.
As in the case of neutral medium, we have for the evolution of $\beta$ with distance,
  \be \label{eq:dbetadr-ion}
    -{d\beta \over dr_{\rm kpc}}=0.01 \times \, {(n_e/{\rm cm}^{-3}) \, r_{\rm kpc} \over (D / 10^{29} \, {\rm cm}^2 \, {\rm s}^{-1})} \,
    \frac{\beta (1-\beta^2)^\frac{3}{2}}{\beta^3+x_m^3} 
     \,.
 \ee
 The resulting relation between $\beta$ and $r$ for a given $\beta_0$ is given by,
 \ba
 && \tan(\arcsin\beta)-(\arcsin \beta)+x_m^3 \ln \tan \Big({\arcsin\beta \over 2}\Big)+\nonumber\\ && \quad \frac{x_m^3}{\cos(\arcsin\beta)}\nonumber\\
 &&=\tan(\arcsin\beta_0)-(\arcsin\beta_0)+x_m^3 \ln \tan \Big({\arcsin \beta_0 \over 2}\Big)\nonumber\\
 && +\frac{x_m^3}{\cos(\arcsin\beta_0)}
 -5\times 10^{-3}\, {(n_e/{\rm cm}^{-3})\, r_{\rm kpc}^2  \over (D / 10^{29} \, {\rm cm}^2 \, {\rm s}^{-1})} \,.
 \label{dbetadr}
 \ea
 The corresponding change in momentum as a function of distance is shown in Figure \ref{fig:momentum} for the case of $\Delta=10$ and $z=10$.

 \subsection{Change in spectrum}
  The loss of energy in protons changes the CR proton spectrum as they diffuse outwards from the virial radius of the parent galaxy.
  The CR proton spectrum  at a given distance $r$ is calculated by using $n_{cr}(\beta (r)) d\beta=n_{cr} (\beta_0) d\beta_0$, which
  follows from the conservation of the number of CRs.
  In order to evaluate it, we use the relation
  between $\beta, \beta_0, r$ from equation \ref{dbetadr-neutral} and \ref{dbetadr}. Note that $n_{cr} (\beta)$ is related to the SFR according to the
  normalisation equation \ref{norm} and has dimensions of time$^{-1}$.
  
\begin{figure}
\centering
\includegraphics[height= 2in]{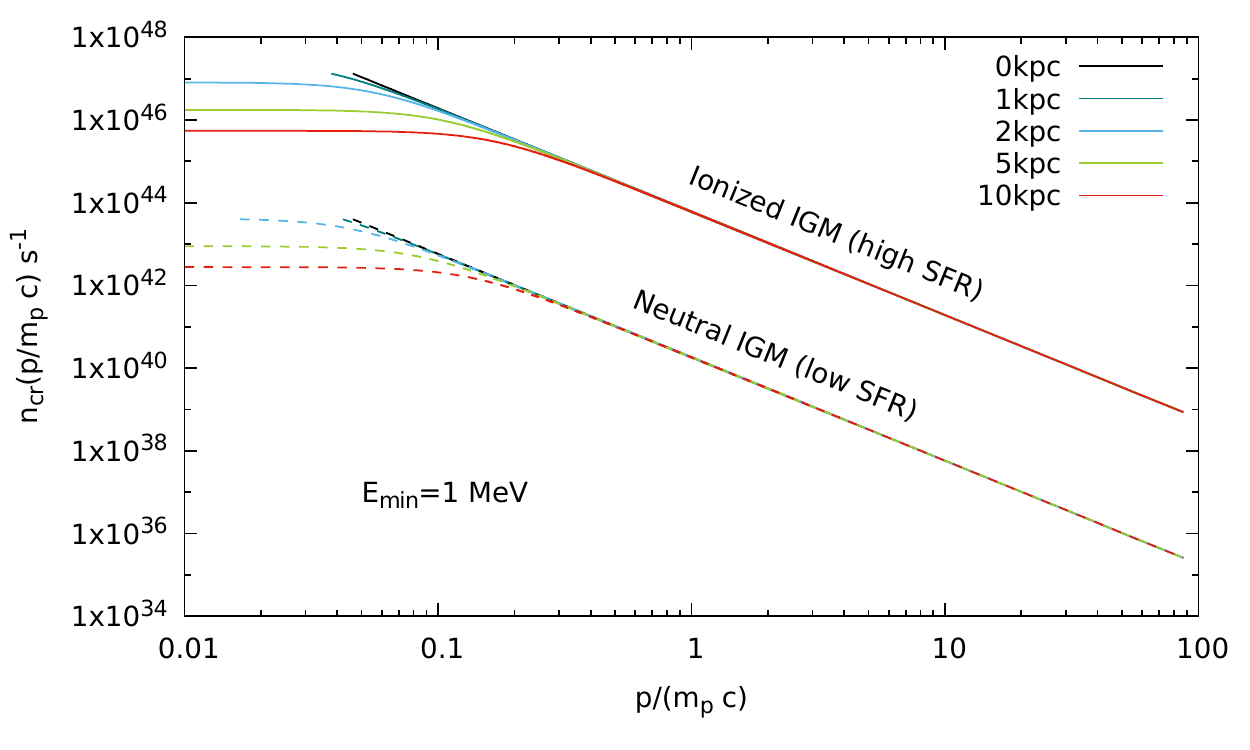}
\caption{The spectrum of CR protons is shown at different distances by solid lines for ionized ($z=10$ and $\rm{SFR=10 M_\odot /yr}$) and by dashed lines for neutral ($z=10$ and $\rm{SFR=0.003 M_\odot /yr}$) medium. Both curves assume $\Delta=10$.}
\label{fig:spectrum}
\end{figure}

 We show in Figure \ref{fig:spectrum} two examples of how the CR spectrum changes at different distances from the galaxy, 
 for the case of SFR=$10$ M$_\odot$ yr$^{-1}$, ionized IGM, (solid curves), and SFR=$0.003$ M$_\odot$ yr$^{-1}$,
 neutral IGM, (dashed curves) both at $z=10$ and assuming $\Delta=10$. As expected, we find that more and more low energy CR protons are depleted as they diffuse outward.

 \section{Gas heating}
 In the case of ionized IGM, the total energy lost by CR protons goes into heating the IGM gas. However, in the case of neutral IGM, only a fraction $f_{\rm heat}$ of the energy lost by protons is used for the heating of the IGM gas, and the rest is spent in partially ionizing the neutral gas and excitation of neutral atoms.
 This  fraction depends not only on the CR proton energy but also on secondary ionization process (by the ejected electrons). Effectively, it depends on the fractional ionization $x_e$ of the gas. As discussed in \cite{Sazonov2015}, this fraction  $f_{\rm heat} \sim 0.25$ for $x_e\sim 0.01$. As a conservative estimate, we use a fraction of $f_{\rm heat}=0.25$ for the case of neutral IGM  and $f_{\rm heat}=1$ for ionized IGM in our calculations.
 
  \begin{figure*}
\centering
\includegraphics[width=7in,height= 4in]{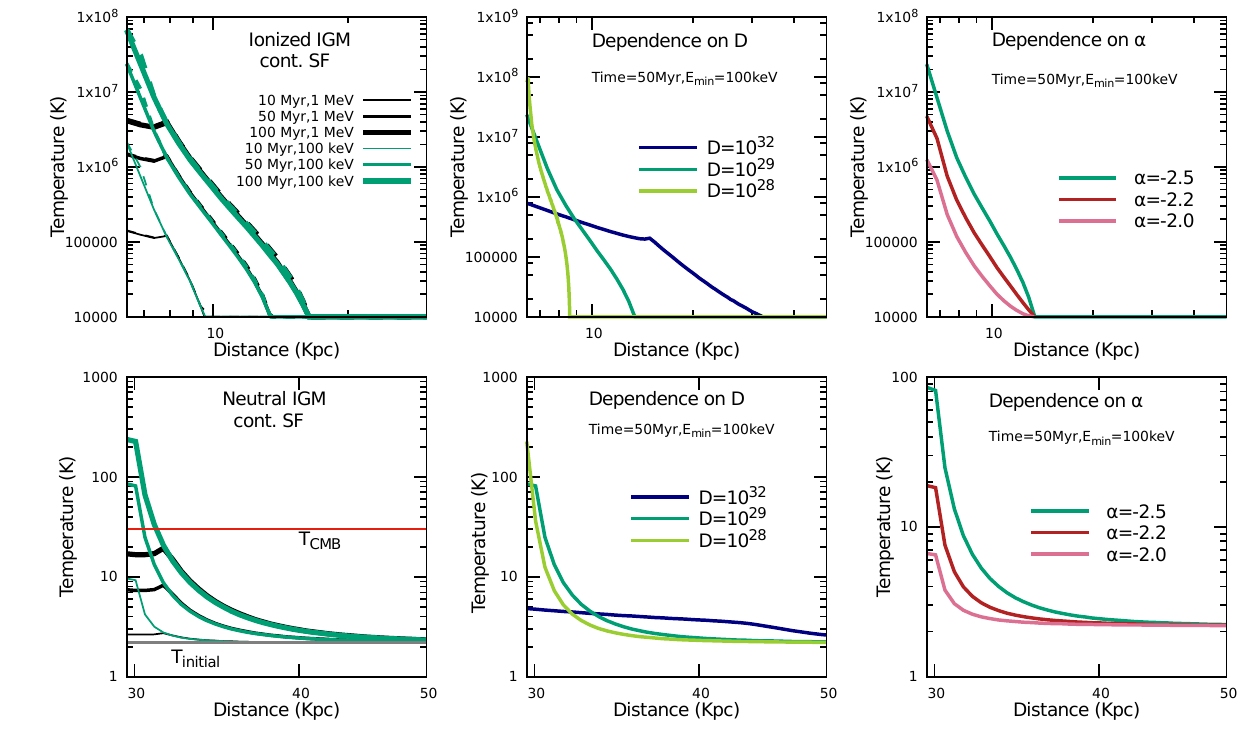}
\caption{The temperature profile for SFR=10$\rm{M_\odot /yr}$, $\rm M_h=10^{10} M_\odot$ (upper panel) and SFR=0.003$\rm{M_\odot /yr}$, $\rm M_h=10^{12} M_\odot$ (lower panel) is shown for $10,50,100$ Myr at $z=10$ using $D=10^{29} \rm cm^2 s^{-1}$ and $\Delta=10$. The left panels shows the case of heating of photoionized gas (upper left) and neutral IGM gas (lower left). Dashed lines show the
profiles without cooling and solid lines show the profiles with cooling. In the lower panels, the horizontal red line corresponds to the CMB temperature.
The middle panels show the variation of the result with diffusion coefficient and the right panels show the variation with CR spectral index $\alpha$.}
\label{fig:temp-profile}
\end{figure*} 

 The rate of increase of energy density $\epsilon$ in a spherical shell of IGM gas
 at distance $r$ and width $\Delta r$ due to interaction of ionized gas with a CR proton of velocity $\beta c$ for a time interval of $\Delta t$
 can be written as,
 \be
   \frac{d\epsilon}{dt}= f_{\rm heat} \, \frac{dE (\beta)}{dt} \, {1 \over 4 \pi r^2 \Delta r}\, {\Delta t}\,.
 \ee
 Here $\Delta t$ is the residence time of a proton in this particular shell during its outward diffusion. We write this as,
 \be
  \frac{d\epsilon}{dt}= f_{\rm heat} \, \frac{dE (\beta)}{dt} \, {1 \over 4 \pi r^2 {d r \over dt}} \, H[t-r^2/(6D)]\,,
 \ee
 where ${dr \over dt}$ refers to the diffusion equation \ref{diffusion}, and the Heavyside step function uses the arrival time ($=r^2/(6D)$) of protons
 at the particular shell at distance $r$. For an ensemble of CR protons, we integrate this over the
 CR spectrum at this shell, $n_{cr} (\beta) d\beta$.
 We finally arrive at,
 \ba
  \frac{d\epsilon}{dt} &=&5\times10^{-19} \, {{\rm erg} \over {\rm s}} \, f_{\rm heat} \, \Big({n_e \over {\rm cm}^{-3}}\Big)\, \frac{1}{4 \pi r^2}\sqrt{\frac{2t}{3D}} \nonumber\\
  && \times \, H[t-r^2/(6D)]\, \int_{\beta_{0min}}^{\beta_{0max}}\frac{\beta^2 n_{cr}(\beta_0)}{\beta^3+x_m^3} d\beta_0  \,.
  \label{depsilondt}
 \ea

 Here we have written the CR spectrum in terms of the initial spectrum $n_{cr} (\beta_0) d\beta_0 (\equiv n_{cr} (\beta) d\beta)$
  in order to explicitly show the limits in terms of the initial values, 
 whose constraints have been discussed in \S 1.
 
 In the approximation of static gas, the energy deposited by CR protons into the gas results in the change in temperature as,
 $d\epsilon={3 \over 2} n_{\rm IGM} k\, dT$.  However, we also take gas cooling into account in order to calculate the change in  temperature 
 with time.

 \subsection{Continuous SF case}
 We first discuss the case of continuous star formation.
 The process of star formation is likely to last for as long as there is gas available. The typical star formation time scale is the inverse 
 of the specific SFR (sSFR), and it decreases from $\sim 10$ Gyr at $z=0$ to $\sim 0.3$ Gyr at $z\ge 2$ \citep{Lehnert2015}. For a
 conservative estimate it is reasonable to assume that  CR heating continues for a time period of $\sim 0.1$ Gyr.
 
 The resulting temperature profiles are shown in Figure \ref{fig:temp-profile} for $z=10$ for two cases: high SFR ($10$ M$_\odot$ yr$^{-1}$ in a $10^{10} \rm M_\odot$ galaxy) with photoionized IGM in the upper left panel, and low SFR ($0.003$ M$_\odot$ yr$^{-1}$ in a $10^{12} \rm M_\odot$ galaxy) with neutral IGM in the lower left panel. Dashed lines show the temperature profile without cooling and solid curves show the profile with cooling, for three different epochs, at $10,50,100$ Myr after the onset of star formation. Black curves show the result of heating with initial energy lower limit of protons at $1$ MeV, and green curves show the profiles when the lower limit is $100$ keV.
 
\begin{figure*}
\centering
\includegraphics[width=5in,height= 3in]{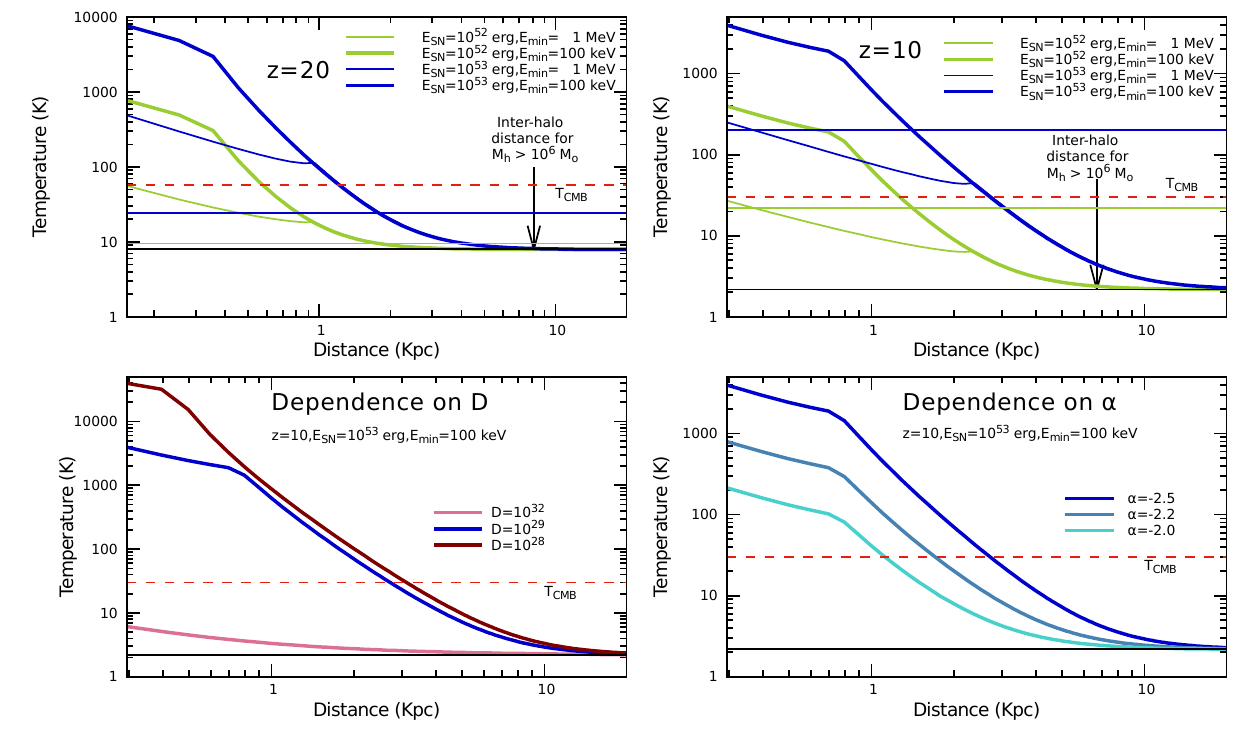}
\caption{The temperature profile for a primordial SN in a mini-halo of $M_h=10^6$ M$_\odot$ at $z=20$ (left-most panel) and $z=10$ (second from left panel). The green curves in both panels show the case for $E_{\rm SN}=10^{52}$ erg and the blue curves, for $E_{\rm SN}=10^{53}$ erg; upper curves are for $E_{\rm min}=100$ keV and lower curves are for $E_{\rm min}=1$ MeV. The horizontal green and blue lines correspond to the increased global mean temperature in each case. The horizontal red dashed lines correspond to the CMB temperature. Vertical arrows mark the inter-minihalo distance for a minimum mass of $10^6$ M$_\odot$. The two panels on the bottom show the variation of the result with diffusion coefficient (bottom left) and CR spectral index $\alpha$ (bottom right), as in Figure \ref{fig:temp-profile}.}
\label{fig:temp-profile-burst}
\end{figure*}

 The profiles show a discontinuity, which stems from the assumption of the initial spectrum being a power law down to a certain minimum energy and zero below it. The discontinuity in the temperature profile occurs at $r_0$ (which is given by equation \ref{eq:rnot})from the virial radius ($\rm R_{vir}=6.4$ kpc for $10^{10} \rm M_\odot$ galaxy and $\rm R_{vir}= 29.6$ kpc for $10^{12} \rm M_\odot$ galaxy) of the source galaxy where the minimum energy proton loses all its energy, as defined earlier. In reality, the spectrum
 will have a smooth change of slope below the minimum energy assumed here, and the temperature profile will consequently be more continuous than shown
 here. However, it is useful to define the distance $r_0$, as we have done here, which indicates a change of shape in the temperature profile.

In the limit of static gas the change in temperature is independent of the density, since $d\epsilon (=1.5 n_{\rm IGM} k \, dT) \propto n_e$, in equation
\ref{depsilondt}. However the profile strongly depends on $\Delta$ since $r_0$ depends on gas density.
 
 As expected from the value of the diffusion coefficient, the heating effect is noticeable only within a few kpc. This is further reduced to when gas cooling is considered. 
 We also show the variation of the result with diffusion coefficient in the middle panels, for three values of $D$. As expected from previous discussion,the local heating decreases with increasing value of $D$. We also show the variation of the results with the CR spectral index $\alpha$, and find that a flatter energy spectrum decreases the heating effect.

\subsection{Minihalo-SF burst case}
Next we discuss the case of a burst of star formation in high redshift minihalos, as considered by \cite{Sazonov2015}. As representative cases, we consider
a minihalo of total halo mass $M_h=10^6$ M$_\odot$ at two redshifts $z=10$, and $z=20$. The corresponding virial radii of the galaxy at these redshifts
are $r_{\rm vir}=0.29,0.15$ kpc. Following \cite{Sazonov2015}, we assume that the average supernova (SN) explosion energy is $E_{\rm SN}=10^{52}\hbox{--}10^{53}$ erg and the average number of 
SNe per minihalo is $f_{\rm SN}=1$. As in the previous section, we assume that a fraction $\eta=0.1$ of the total SNe energy is converted into accelerating CR 
particles.

We can adopt the gas heating equation \ref{depsilondt} to the case of a burst of CR particles, by writing  $n_{cr} (\beta_0)= N_{\rm cr} (\beta_0) \delta(t- [r^2/(6D)])$ within the integral. The distribution function of CR particles produced in the burst is normalised by,
\be
\int E_k \, N_{\rm cr} (p_0)\, dp_0 = \eta f_{\rm SN} E_{\rm SN} \,,
\ee
which is similar to equation \ref{norm2}.

In the absence of cooling in the neutral IGM (since the resulting temperature change is shown to be  small below), the CR particles heat up the surrounding as
they diffuse and sweep past the IGM gas. The change in the energy density of gas at distance $r$ at time $t$ due to CR protons with initial $\beta_0$ is given by,
\ba
\Delta \epsilon (r) &=& \int_0 ^t n_{cr} (\beta_0) f_{\rm heat} {dE \over dt} {1 \over 4 \pi r^2} \sqrt{{2t \over 3D}} \, dt  \nonumber\\
&=& {1 \over 4 \pi r^2} \int_0 ^t N_{\rm cr} (\beta_0) \delta [t-{r^2 \over 6D}] \, f_{\rm heat} {dE \over dt} \sqrt{{2t \over 3D}} \, dt \nonumber\\
&=&{2.9 \times 10^{-19} \over 12 \pi D \, r} \Bigl ( {n_{\rm HI} \over {\rm cm}^{-3}} \Bigr ) f_{\rm heat}{2 \beta^2 \over \beta_c^3 + 2 \beta^3} \, N_{\rm cr} (\beta_0) \,.
\ea
Here ${dE \over dt}$ refers to energy loss of a CR proton in  neutral medium (equation \ref{eq:energyloss-neutral}).

The resulting temperature difference at distance $r$ from the minihalo is found by a simple integration over the energy spectrum of CR to be,
\be
\Delta T = {7.4 \times 10^{-5} f_{\rm heat} \over D \times r} \times \frac{n_{\rm HI}}{n_{\rm IGM}}  \int_{\beta_{0min}}^{\beta_{0max}}\frac{\beta^2 N_{\rm cr}(\beta_0)}{2\beta^3+\beta_c^3} d\beta_0  \,.
\label{eq:deltaT}
\ee
Again, the temperature profile is dependent on the assumption of $\Delta$ through the relation between $\beta_0$
 and $\beta$ by equation \ref{dbetadr-neutral}.We show the results in Figure \ref{fig:temp-profile-burst} for the case of $E_{\rm SN}=10^{52}$ erg for PopIII stars and $f_{\rm SN}=1$, at two redshifts $z=10, 20$. 
The more optimistic case of a pair instability SN with $E_{\rm SN}=10^{53}$ erg is shown by the upper curve, for which the value of $\Delta T$ is an order of magnitude larger than the fiducial case, and the IGM temperature exceeds the CMB temperature near the virial radii. 

The integral in the above equation is roughly constant up to a distance of $r_0$ from the virial radius,  and decreases as $r^{0.7 \alpha}$ beyond that, roughly up to $\sim 2 r_0$. Therefore the temperature profile for $0.1 \rm MeV<E_{\rm min}<1 MeV$ can be written as,\\
\ba
\Delta T & \approx & 25\, {\rm K} \, \Bigl ( {f_{\rm heat} \over 0.25} \Bigr )\,\Bigl (  {\eta \,f_{\rm SN} \,E_{\rm SN} \over 0.1 \times 10^{52} \, {\rm erg}} \Bigr ) \, \Bigl ({E_{0,min} \over 1\, {\rm MeV}} \Bigr )^{-1.2}\times \nonumber\\ && \Big(\frac{|\alpha|}{2.5}\Big)^{10.8\times \Big(\frac{E_{0,min}}{\rm 1 MeV}\Big)^{-0.1}} \times (r_{\rm kpc}+ R_{\rm vir})^{-1} \times \nonumber\\ && \Bigl ({D \over 10^{29} \, {\rm cm}^2 \, {\rm s}^{-1}} \Bigr )^{-1} 
 \, \qquad ,R_{\rm vir}<r<(r_0+R_{\rm vir}) \nonumber\\
\Delta T&\approx& \Delta T(r_0+R_{\rm vir})\Bigl({r_{\rm kpc} + R_{\rm vir}\over r_0+R_{\rm vir}} \Bigr )^{[0.7\alpha-1]}\nonumber\\&& \qquad \qquad \qquad ,(r_0+R_{\rm vir})<r<(2r_0+R_{\rm vir})\,
\label{eq:deltaTfit}
\ea 
If the diffusion coefficient increases beyond the fiducial 
value at high redshift, because of a lower magnitude of magnetic field in the IGM, then the temperature profile extends to larger distances but with a lower
magnitude.

Figure \ref{fig:temp-profile-burst} also shows the typical distances between mini-halos ($R_{\rm interhalo}$), assuming a minimum halo mass of $10^6$ M$_\odot$ with 
vertical arrows. We can then use our calculated temperature profile to determine the average increase in temperature within this length-scale, 
given by,
\be
\Delta T_{\rm avg}={\int \Delta T 4 \pi r^2 dr \over (4/3)\pi (R_{\rm interhalo}^3-R_{\rm vir}^3)}\,.
\label{deltaT_local}
\ee

This value should be compared with the global temperature increase calculated by \cite{Sazonov2015}. Following them, if we  define a fraction $\eta_{\rm LECR}$ as the product of CR acceleration efficiency $(\eta)$ and the energy fraction carried by low energy CRs (which deposit their energy into the IGM within the Hubble time), and $n_h (z) $ as the number density of mini-halos, then according to their equation 11, the global increase in temperature is given by,
\be
 \Delta T _{\rm IGM} ={ f_{\rm heat} \eta_{\rm LECR} f_{\rm SN} E_{\rm SN}\over (3/2) k \, n_{\rm IGM}(z)} n_h (z) \,
 \label{deltaT_global}
\ee
 We have calculated this value using the appropriate $n_h(z)$, for $M_{\rm min}=10^6$ M$_\odot$, $M_{\rm max}=10^7$ M$_\odot$, by using the {\it CAMB} transfer function calculator and the fitting function of \citet{reed2007}. The calculation has been done by the HMF calculator given by \citet{murray}.
The fraction $\eta_{\rm LECR}$ depends on the assumed CR spectrum, and we use the appropriate values in our calculation. For $\alpha=-2.5$, the energy fraction in low energy cosmic rays ($\le 30$ MeV) is $0.17$, and for $\alpha=-2.2$, it is $0.05$. Since we have used a cosmic ray acceleration efficiency($\eta$) of $10\%$, we have $\eta_{\rm LECR}=0.017$ and $0.005$ for $\alpha=-2.5$ and $-2.2$, respectively.

If the local average as calculated using equation \ref{deltaT_local} exceeds the global average increase in temperature, then it would
imply that the heating by CR is patchy, and the temperature profiles presented here are representative of the effect of CR heating. On the other hand, if the local average is less than the global average increase in temperature, then it would mean that CR heating is rather uniform and the temperature profile calculated by us would be subsumed under the global increase in temperature.

We find  from equation \ref{deltaT_global} that for $E_{\rm SN}=10^{53} \rm erg$, the global temperature increase is 198 K at redshift 10 and 16 K at redshift 20, whereas the local average temperature increase from equation \ref{deltaT_local} is 16 K and 2 K at redshift 10 and 20 respectively. Therefore we can put upper bounds on the diffusion coefficient $D$ at high redshift, for which CR heating would be inhomogeneous (larger $D$
would imply a more uniform heating). We have found that at $z=10$, the limit is $D \le 1\times 10^{26}$ cm$^2$ s$^{-1}$.
 Therefore for our fiducial value (see \S 2.2) of $D\sim 10^{29}$ cm$^2$ s$^{-1}$, the heating is likely to be uniform. At $z\sim 20$, the corresponding limit is $D\le 5\hbox{--}6\times 10^{26}$ cm$^2$ s$^{-1}$, which also implies uniform heating since $D$ is likely to be above this limit. However, one should remember there are uncertainties in the evolution of magnetic field with redshift and the dependence of $D$ on the magnetic field.

\subsection{Effect of a density profile}
Having calculated the temperature profile using a constant IGM density, we now show  the effect of a density profile, by
assuming a simple power-law relation. Simulations of accretion of mass around massive halos have shown that the density
profile around the virial radius of haloes is steeper than $r^{-2}$, but becomes flatter than $r^{-2}$  beyond the virial radius (up to a distance of $\sim 5 R_{\rm vir}$) \citep{Prada2006}. Moreover, the overdensity at $\sim R_{\rm vir}$ is $ \Delta\sim100$. The overall profile from $R_{\rm vir}$ to $\sim 5R_{\rm vir}$ can therefore be approximated by,
\be
n(r)=100 {\rho_{\rm cr}(z) \Omega_m (z) \, f_b \over \mu m_p} \Bigl ( {r \over R_{\rm vir}} \Bigr )^{-2} \,.
\label{eq:density-profile}
\ee
This profile would change the $r$-dependence in the equations relating $\beta$ and $\beta_0$ (equations \ref{dbetadr-neutral} 
and \ref{dbetadr}) that can be analytically calculated.

\begin{figure}
\centering
\includegraphics[height= 2in]{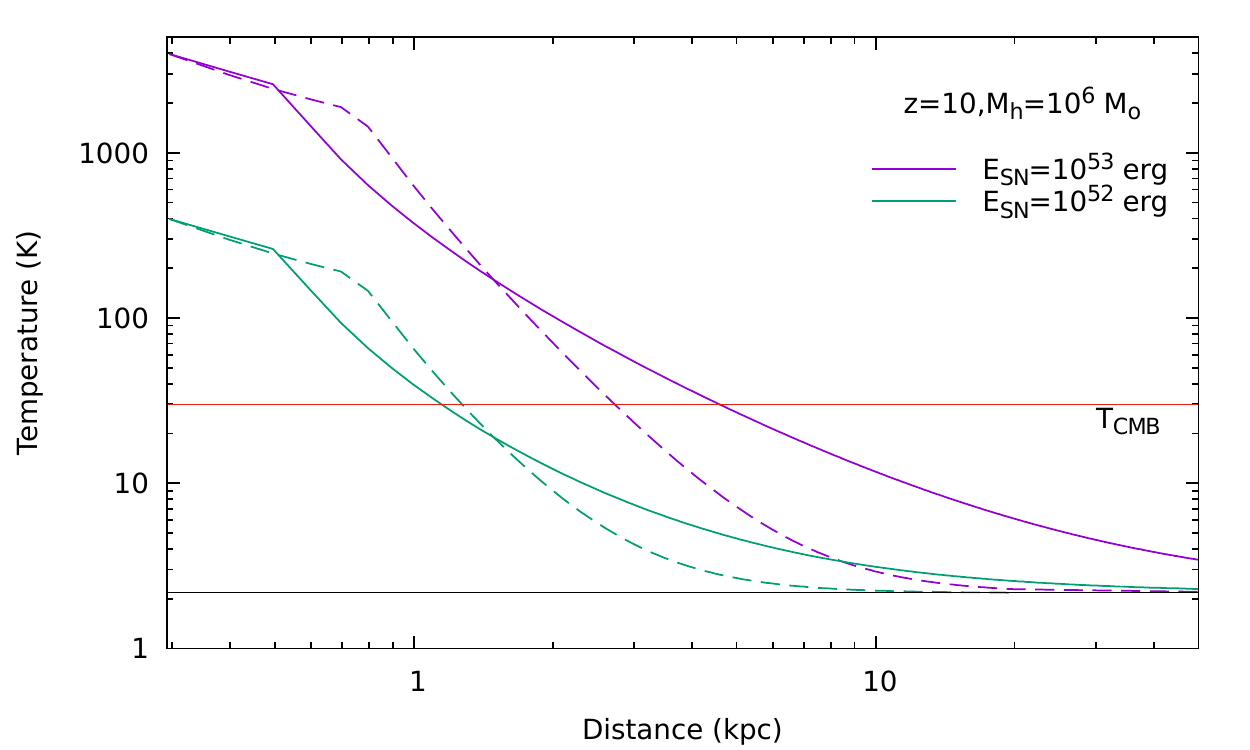}
\caption{The temperature profile for a primordial SN in a mini-halo of mass $10^6 M_\odot$ and  at $z=10$ assuming $\rm E_{min}=0.1 MeV$. The dashed curves show the temperature profiles when the gas density outside the halo is 10 times the baryonic matter density at that epoch and the solid curves are obtained using a density profile of gas around the halo. }
\label{fig:density-profile1}
\end{figure}

We show the change in the temperature profile for the mini-halo case in Figure \ref{fig:density-profile1}, for $z=10$, and $M_h=10^6$ M$_\odot$, for two values of $E_{\rm SN}$ and $E_{\rm min}=0.1$ MeV. The dashed profiles correspond to a uniform density with $\Delta =10$ as assumed earlier (see Figure \ref{fig:temp-profile-burst}). The solid lines show the case of the above mentioned density profile. The approximate concurrence of these two curves justifies our assumption of $\Delta=10$ for the uniform density case.

\begin{figure}
\centering
\includegraphics[height= 2in]{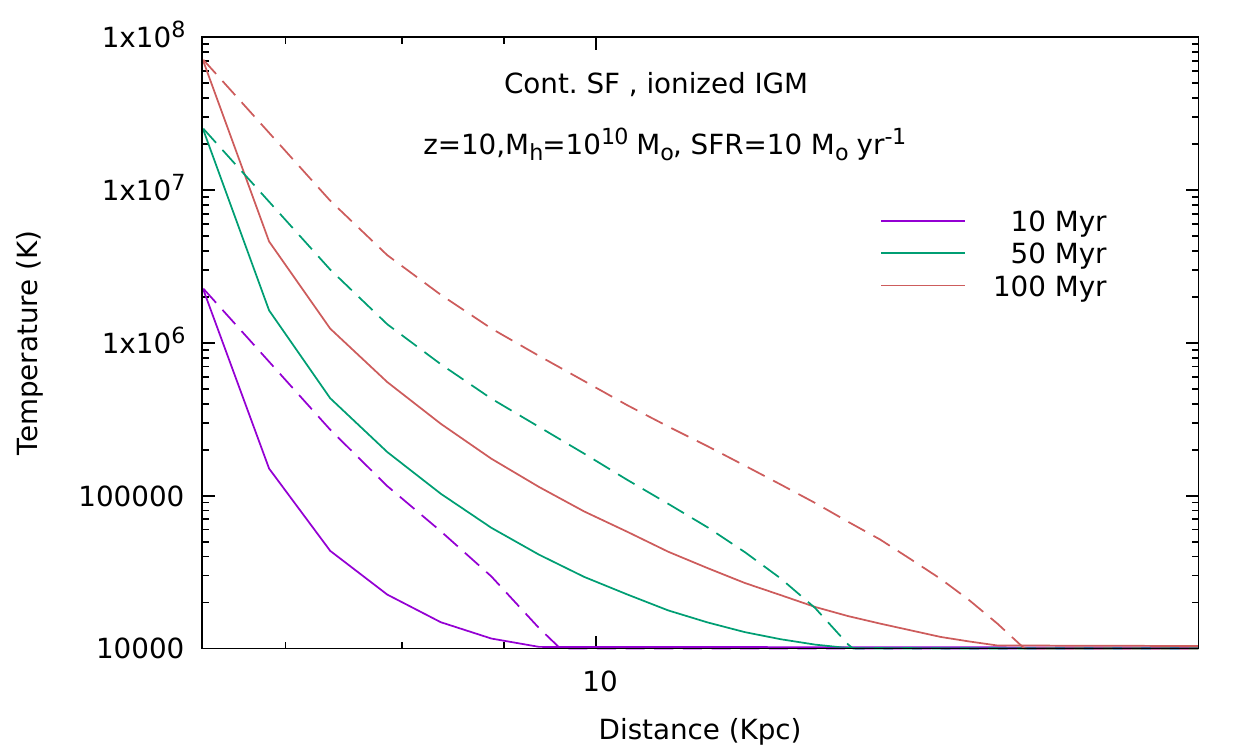}
\caption{The temperature profile for continuous star formation in a galaxy of $\rm M_h=10^{10} M_\odot$ at $z=10$ assuming $\rm E_{min}=0.1$ MeV. The dashed curves show the temperature profiles when the gas density outside the halo is uniform and 10 times the baryonic matter density at that epoch and the solid curves are obtained using a density profile of gas around the halo as in equation \ref{eq:density-profile}. }

\label{fig:density-profile2}
\end{figure}

The case of a continuously star forming galaxy with a density profile outside $R_{\rm vir}$ is shown in Figure \ref{fig:density-profile2}. Here, although the temperature near the virial radius is similar to the case of uniform density (again justifiying the assumption of $\Delta=10$), the temperature profile is steeper than the uniform density case, because of the change in the relation between $\beta$ with distance.

 \section{Discussion}
 The temperature profile in the neutral case  shows that if the SFR is to be low enough to keep the surrounding gas
 neutral and high enough to cause substantial heating, as in the case portrayed in the lower panel of Figure \ref{fig:temp-profile}, the gas temperature
 can exceed the CMB temperature (shown by a horizontal red line) within a few kpc of the galaxy, if the lower limit of CR proton energy is $100$ keV.
 This was the scenario sketched by \cite{Sazonov2015}, which we have quantified here, and shown the dependencies on various parameters. 
 
 However, even if such a case of heating arises, it is unlikely to be probed in the near future by observations as the corresponding angular scale
 is very small, of the order of a few arc seconds. 
 
\begin{figure}
\centering
\includegraphics[height= 2in]{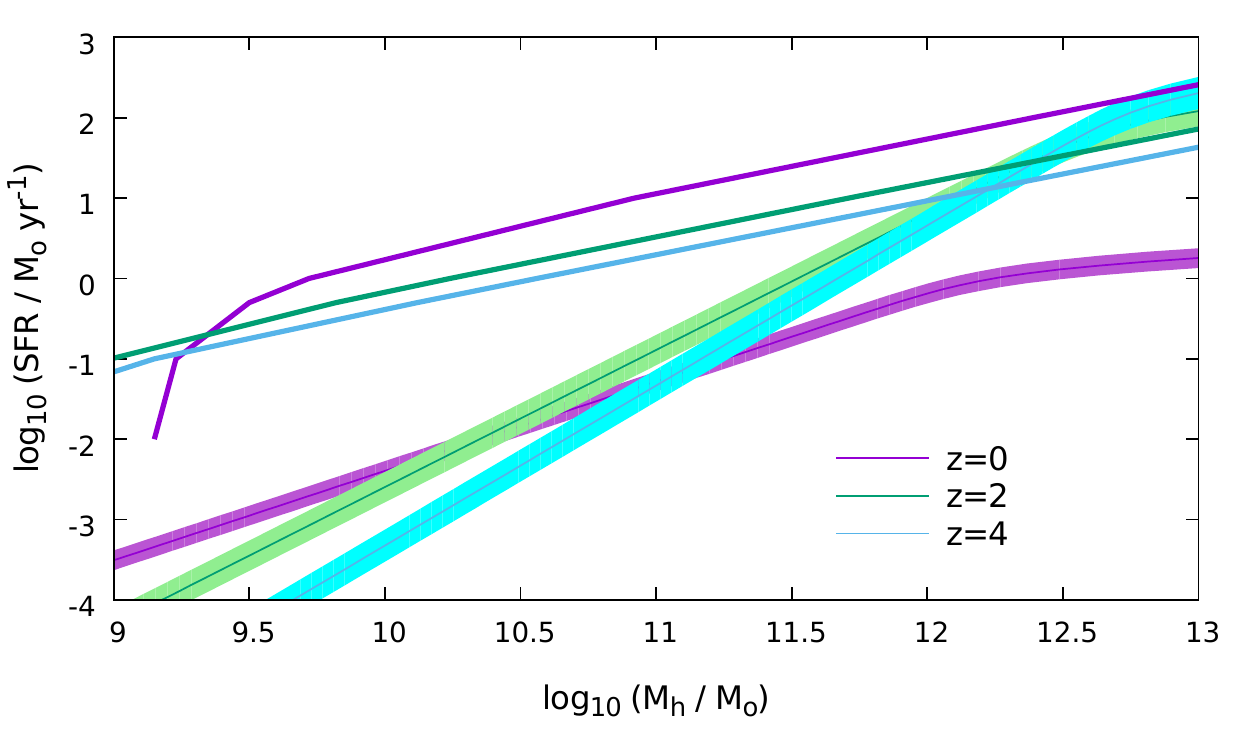}
\fontsize{13}{12}
\caption{SFR for different galaxies masses are shown for $z=0,2,4$. Shaded lines show the star formation rates of main sequence of galaxies, and the solid lines show the lower limits of SFR from the condition that sound speed of the heated gas near the virial radius should exceed the infall speed, $E_{0,min}=100$ keV.}
\label{fig:SFR}
\end{figure}
 
 The temperature profiles in Figure \ref{fig:temp-profile} show that gas temperature in the ionized case can increase to $10^7$ ($10^6$) K for $E_{0,min}=0.1 (1)$ MeV
 in a time period of $\sim 50 $ Myr, for SFR of $10$ M$_\odot$ yr$^{-1}$. This implies that the gas in the outskirts of the galaxy likely to be heated up to a high temperature, which would
 set up an outward motion. This will affect the gas infall into the parent galaxy and in turn influence the evolution of the star formation process in it.
 
 It is reasonable to argue that if the sound speed of the gas heated by CR exceeds the infall velocity near a galaxy, then CR heating will
tend to suppress the further gas infall. The infall velocity is roughly estimated as $\sqrt{GM_h/R_{vir}}$, and independent of the distance from the 
galaxy,  up to $\sim 1.5 \times R_{\rm vir}$ \citep{Goerdt2015}. Therefore we can determine the minimum SFR needed to inhibit gas infall around a galaxy of a given mass at a certain redshift. 
We can then compare it to the
SFR of main sequence of galaxies appropriate for galaxies of the same mass at that redshift.

We show in Figure \ref{fig:SFR} the SFR of main sequence of galaxies as a function of halo mass at $z=0,2,4$ by shaded lines. We have used the fit to SFR as a function of stellar mass and cosmic time as given by \cite{Speagle2014}, and the analytical fit for the relation between stellar mass
and halo mass, as given by \cite{Behroozi2010}, and the shaded region show $1-\sigma$ error bars. 
Superposed in the same figure are lines that show the lower limits on SFR needed to suppress
gas infall by CR heating near the virial radius, for $E_{0,min}=100$ keV (solid lines).
For the calculation at these low redshifts, we have used a 
fiducial value $D=10^{29} \rm cm^2 s^{-1}$ as discussed in \S 2.2.

We find that, if the lower limit of CR protons  is $100$ keV when they escape from galaxy, then CR heating affects the infall of gas around galaxies of
M$_h\sim 10^{10}$ M$_\odot$ for SFR  of order $\sim 1$ M$_\odot$ yr$^{-1}$. 
For more massive galaxies, CR heating can be important for suppression of infall if the SFR $\ge 100$ times the SFR of main sequence of galaxies. at $z=0$.  This gap (between required SFR and SFR of main sequence of galaxies) narrows with increasing redshift. At $z\sim 4$, infall around galaxies
with $M_h \sim 10^{12}$ M$_\odot$ can be affected for the SFR of main sequence of galaxies. 
Therefore CR heating can be an important feedback mechanism for regulating gas infall around massive galaxies at high redshift.

\section{Summary}
We have calculated the radial temperature profile of the IGM around galaxies due to heating by CR protons after taking into
account the effect of diffusion of CRs. We assumed a simple power-law CR spectrum with a low energy cutoff, and a constant
diffusion coefficient for low energy CRs. 
We considered three cases: (1) heating of neutral IGM at high redshift around galaxies
with low enough SFR so that the IGM is not photoionized, (2) heating of photoionized IGM around high SFR galaxies and (3) heating of
neutral IGM at high redshift around a minihalo on account of  primordial supernovae.
 Our main results are:
\begin{itemize}
\item It is not easy for low energy CRs to escape relatively massive galaxy (Figure \ref{fig:grammage}).
\item The surrounding medium of galaxies at high redshift is likely to be ionized on account of star formation in the galaxy. Therefore,
the heating by CRs will proceed in a different manner than previously considered case of neutral IGM heating.
\item In the case of CRs from mini-halos at $z\sim 10\hbox{--}20$, we put an upper bound on the
diffusion coefficient ($D \le 1 \times 10^{26}$ cm$^2$ s$^{-1}$ for $z\sim10$ and 
$D\le 5\hbox{--}6 \times 10^{26}$ cm$^2$ s$^{-1}$ for $z\sim 20$) for which the heating is inhomogeneous,
after comparing our temperature profiles with the estimate of global temperature increase. Given the expected scaling of 
the diffusion coefficient with redshift, this bound suggests uniform heating, both at $z\sim 10$ and 20. 
But the uncertainties in CR parameters (spectrum, lower energy limit of emerging CRs, diffusion coefficient and its dependence on magnetic field) and magnetic field at high redshift precludes any 
firm conclusion.
\item In the case of continuous star formation and neutral IGM in the vicinity of galaxies with low SFR, the temperature exceeds the 
CMB temperature for $D$ $< 10^{30}$ cm$^2$ s$^{-1}$, and in this case the profile is too peaked to be detectable.
\item Furthermore, we found that the heated gas
near the virial radii of galaxies can provide a feedback mechanism by inhibiting the infall of gas, especially for massive galaxies at 
high redshift ($z\sim 4$), and low mass star forming galaxies at low redshifts for sufficiently high SFR.
\end{itemize}

\section*{Acknowledgements}
We would like to thank  an anonymous referee for invaluable comments and P. L. Biermann, S. Biswas, P. Sharma and Y. Shchekinov for valuable discussions.
\footnotesize{

\end{document}